\documentclass[sigconf]{acmart} 

\usepackage{graphicx}
\usepackage{tabularx}
\usepackage{multirow}
\usepackage{enumitem}
\usepackage{comment}
\usepackage{soul}
\usepackage{todonotes}
\usepackage{enumitem}
\usepackage{placeins}
\usepackage{csquotes}
\usepackage{tikz}
\usepackage{bbm}

\usepackage{soul}
\usepackage{xspace}

\setlist[description]{leftmargin=\parindent,labelindent=0pt,itemsep=0pt}

\definecolor{darkgreen}{rgb}{0.0, 0.5, 0.0}

\definecolor{anne}{rgb}{0,0.5,0.9}
\definecolor{timo}{rgb}{0.998,0.722,0.635}
\definecolor{lukas}{rgb}{0,0.5,0.9}
\definecolor{simon}{rgb}{0.998,0.722,0.635}

\usepackage{utfsym}

\usepackage{csquotes}

\AtBeginDocument{%
  \providecommand\BibTeX{{%
    Bib\TeX}}}

\begin{document}

\title[AI-Based Feedback in Counselling Competence Training of Prospective Teachers]{AI-Based Feedback in Counselling Competence\\ Training of Prospective Teachers}

\author{Tobias Hallmen}
\orcid{0009-0005-6450-5694}
\affiliation{%
  \institution{Chair for Human-Centered \\ Artificial Intelligence}
  \city{University of Augsburg}
  \country{Germany}
}
\email{tobias.hallmen@uni-a.de}

\author{Kathrin Gietl}
\orcid{0009-0004-7584-2226}
\author{Karoline Hillesheim}
\orcid{0009-0005-8609-2122}
\affiliation{%
  \institution{Chair of Primary School Pedagogy \\ and Primary School Didactics}
  \city{University of Augsburg}
  \country{Germany}
}
\email{kathrin1.gietl@uni-a.de}
\email{karoline.hillesheim@uni-a.de}

\author{Moritz Bauermann}
\orcid{0009-0000-5786-0149}
\affiliation{%
  \institution{Department of \\ Medical Education Augsburg}
  \city{University of Augsburg}
  \country{Germany}
}
\email{moritz.bauermann@uni-a.de}

\author{Annemarie Friedrich}
\orcid{0000-0001-8771-7634}
\affiliation{%
  \institution{Professorship Natural Language \\ Understanding with Applications to Digital Humanities}
  \city{University of Augsburg}
  \country{Germany}
}
\email{annemarie.friedrich@uni-a.de}

\author{Elisabeth Andr\'e}
\orcid{0000-0002-2367-162X}
\affiliation{%
  \institution{Chair for Human-Centered \\ Artificial Intelligence}
  \city{University of Augsburg}
  \country{Germany}
}
\email{elisabeth.andre@uni-a.de}

\renewcommand{\shortauthors}{Hallmen et al.}

\begin{abstract}
This study explores the use of AI-based feedback to enhance the counselling competence of prospective teachers. An iterative block seminar was designed, incorporating theoretical foundations, practical applications, and AI tools for analysing verbal, paraverbal, and nonverbal communication. The seminar included recorded simulated teacher-parent conversations, followed by AI-based feedback and qualitative interviews with students. The study investigated correlations between communication characteristics and conversation quality, student perceptions of AI-based feedback, and the training of AI models to identify conversation phases and techniques. Results indicated significant correlations between nonverbal and paraverbal features and conversation quality, and students positively perceived the AI feedback. The findings suggest that AI-based feedback can provide objective, actionable insights to improve teacher training programs. Future work will focus on refining verbal skill annotations, expanding the dataset, and exploring additional features to enhance the feedback system.

\end{abstract}


\begin{teaserfigure}
  \centering
  \includegraphics[width=0.75\textwidth]{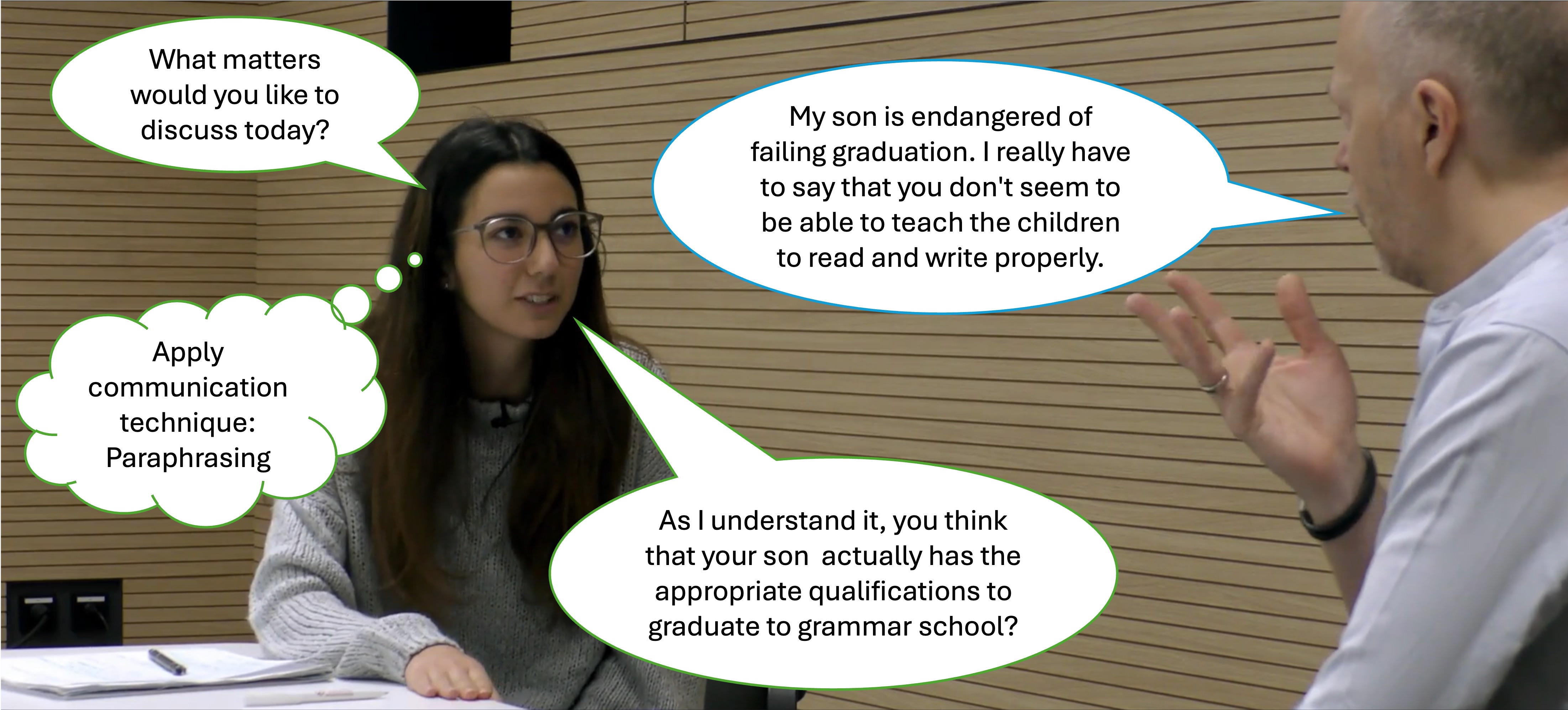}
  \caption{Simulated teacher-parent conversation.}
  \label{fig:teaser}
\end{teaserfigure}

\received{20 February 2007}
\received[revised]{12 March 2009}
\received[accepted]{5 June 2009}

\maketitle


\section{Introduction}

Advising parents on their children at school is one of the most stressful tasks for teachers according to \citet{landert2009lch}. Teacher-parent conversations are often unprofessional (\citet{sauer2015beraten, wegner2016lehrkraft}), unsatisfactory, exhausting, and time-consuming for both parents and teachers (\citet{aich2019gesprachsfuhrung, sacher2005erfolgreiche}). But why keep having these conversations, if both parties dislike them deeply?

It is shown by \citet{cox2005evidence, epstein2001more} that the cooperation of parents and teachers on school-relevant topics improves pupils' performance, reduces inappropriate behaviour, and is beneficiary for all parties. Therefore, identifying pupils' problems in school during teaching, listening to parents' concerns and wishes, making compromises and developing solutions, and counselling parents towards this goal is one of the core practices of teachers according to \citet{grossman2021teaching}.

In a nutshell, the quality of teacher-parent communication is crucial for the pupils' well-being and success around and in school. Traditional training methods suffer from the raters subjective judgement and humans incapability to measure reliably features, such as speaking rate or loudness. AI models offers a novel solution addressing these gaps by providing objective feedback and making features measurable.

In this work, we focus on the communicative skill (split into three subcategories: verbal, paraverbal, and nonverbal) and address how these competencies can be fostered applying AI models during the study program of prospective teachers in Germany. To this end, we designed an iterative block seminar, where students learn the relevant communication theory and how to apply it in counselling. It starts with the theoretical foundation and becomes more and more practical, culminating in a simulated teacher-parent counselling session at the end. Based on this recorded simulated conversation, the students receive feedback from instructors as well as AI-based feedback. Afterwards, the students are interviewed and questioned on the concept of the seminar, how they perceived the different displays of feedback, and what needs further improvement. These interviews are then analysed qualitatively and used to improve the block seminar for the next semester.

This work investigates: (I1) Correlations between nonverbal and paraverbal characteristics and conversation quality, (I2) Student perceptions of AI-based feedback, and (I3) Training AI models to identify conversation phases and techniques.


Since the students' feedback flows back into the improvement of the next held seminar, thereby introducing variability, we report on the saturated observations.


\section{Background and Related Work} \label{sec:background}
A meta-study by \citet{cook2014much} in medicine showed average positive effects for simulated professional conversations between medical doctors and patients compared to other forms of teaching, e.g. lectures or videos. There was no significant difference between peers and actors as patients, but students preferred actors. \citet{lane2007use} corroborate this by showing that trainings done with actors are rated positively by students. Having these trainings alone is already beneficial, but adding feedback increases their efficacy, according to \citet{gerich2016teachers}. The quality, as rated by experts, of these (simulated) conversations depend on the competence of the teachers in giving professional counsel according to \citet{gerich2015skills}, which comprises the four main skills communication, diagnostic, problem-solving, and coping.

Attempts have been made to employ AI-based tools to automatically analyse conversations between patients and therapists to provide insights on outcome, such as symptom reduction, and process variables, such as affective experiences. Qualitative studies in the context of patient-focused psychotherapy (\citet{baur:et:al:2020}) revealed the promise of AI-based tools for psychotherapy research and practice. \citet{terhuerne:et:al:2022} evaluated the capabilities of the NOVA annotation tool to assess non-verbal expressions in psychotherapy. An interesting finding of their research was that the assessments of NOVA correlated with the therapist's assessment of the patient's emotions, but not with the patient's self-reports, indicating that NOVA is closely related to the external assessment of observable emotions. 

While Baur et al. highlighted the challenges of manual annotation and subjective interpretations, Terhürne et al. focused on improving emotion recognition models using larger datasets and validating their predictive value for symptom severity. Both studies underscore the potential of NOVA in enhancing the analysis of non-verbal behaviors in therapeutic settings, or abstracted -- professional counsellings.

To our knowledge, the use of AI-based annotation tools has not yet been explored in the context of German elementary school teacher-parent conversations. 
In this work, we focus on the communicative part of such professional counsellings, which consists of three subcategories:

(1) Verbal communication is essentially the spoken language and the rules it follows like grammar, syntax, and semantics (\citet{rocci2016verbal}). Here we focus on specific constructs that students are being taught to use in counselling, i.e. conversation phases and communication techniques.

(2) Paraverbal communication is essentially the audible part of communication. Exemplary paraverbal features are pitch, resonance, articulation, loudness, melody, and pauses (\citet{matoba2007paraverbal}).

(3) Nonverbal communication is the part which is neither verbal nor paraverbal (\citet{knapp1972nonverbal}), i.e. unspoken communication. Nonverbal features are for instance directed and averted gaze, smiling, facial emotions, gestures, etc.

To investigate (I1) -- (I3) we designed a four-day block seminar:
\begin{enumerate}
    \item Conversation phases, communication techniques, peer roleplay
    \item counselling competence at school, nonverbal communication, coping strategies
    \item Recorded simulated teacher-parent conversation, reflection and expert feedback
    \item AI-based feedback, interviews
\end{enumerate}

The students' feedback flows back into improving the next block seminar and the AI-based analysis around it, leading to an iterative design and improvement process.

\section{Dataset}
The seminar takes place every semester and at the time of writing, four seminars took place totaling in $n = 29$ students (3 male, $10.34\%$) of elementary school pedagogics, i.e. on average $7.25$ students per semester. They are between 19 and 26 years old (avg = $21.34a$, sd = $2.11a$) and are between the 2nd and 12th semester (avg = $4.69$, sd = $2.73$). The simulations' durations range from $429.27s$ to $865.33s$ (avg = $610.54s$, sd = $105.43s$) and are held with two actors (1 male, $50\%$) playing the role of the parent.

The topics of the simulations are diverse, requiring different solution finding and counselling skills: divergence of parents' wishes on which  secondary school they want their child to attend and for which secondary school the child's grades suffice; reasons and mitigations on why parents do not want their child to participate in school's swimming lessons; finding origins of behavioural abnormality during lectures and ways to address them; educational counselling on the usage of mobile devices and causing distractions at school; learning advice on the lack of homework done or the lack of enthusiasm if doing homework; educational counselling on the effects of parents' divorce and how to reduce them.

For the initial seminar, the first batch of simulations ($n = 6$) were conducted in a recording studio, featuring radio lapel microphones sampling at $48kHz$ and four $2160p25$ cameras: one microphone per interlocutor, one close shot camera per face, one total view camera, and one camera in medium range, showing upper body of the teacher behind their desk, cf. Figure~\ref{fig:teaser}.

The studio is used to dramaturgic shots and cutting, therefore the recordings look and feel good (think of watching a movie), but there are issues: there is spillover in the audio recordings, requiring tedious manual cleanup, the close shot cameras are angled, reducing the accuracy of facial expression feature extraction, the audio/video streams are not synchronized, and the participants had to sit sideways, reducing the naturalness of such a conversation in addition to all the surrounding recording equipment.

To mitigate these issues and to reflect the modern world of school, where parents and teachers can communicate digitally over an app called ``Schulmanager'' (``school manager''), we switched for the remainder of videos ($n = 23$) to Zoom, a video conferencing software, featuring a separate video and audio stream ($16kHz$) per participant. This removed all aforementioned problems, only introducing little delays caused by round trip time. Depending on both terminal devices, the majority of videos is recorded in $360p25$, sometimes in $540p25$, showing the upper bodies of the interlocutor in frontal view.

\section{Methodology}

\subsection{Feature Selection}

For the verbal part we adapt conversational phases as defined by \citet{benien2003schwierige}, cf. Table~\ref{tab:phases} -- beginning, informational, argumentative, decision-making, and concluding phases -- and communication techniques as proposed by \citet{gerich2015skills}, that are extended and operationalized by definitions of \citet{bay2021erfolgreiche} and \citet{gartmeier2018gesprache}, cf. Table~\ref{tab:techniques} -- paraphrasing, verbalising, and structuring . It is a central part of the theory that the students are being taught in the seminar. They are ought to apply it during the simulation and, as part of a self-reflection exercise, annotate their transcript of their simulated conversation with conversation phases and communication techniques. 

\begin{table}
\caption{Conversational phases, their subcategories and short descriptions following \citet{benien2003schwierige}.}
\label{tab:phases}
\begin{tabularx}{\linewidth}{@{}p{2cm}p{2cm}X@{}}
\toprule
\textbf{Phase}  & \textbf{Subphase}       & \textbf{Description}                                                                  \\ \midrule
Beginning       & Greeting                & The interlocutor greet each other.                                                    \\ \cmidrule(l){2-3} 
                & Smalltalk               & Lightweight conversation without going into depth.                                    \\ \cmidrule(l){2-3} 
                & Time Frame              & A time is specified how much time is available for the conversation.                  \\ \cmidrule(l){2-3} 
                & Content Frame           & The topic of the conversation is clarified and participants formulate their concerns. \\ \midrule
Informational   & --                      & Information on the situation is exchanged without developing solutions.               \\ \midrule
Argumentative   & --                      & Discussing suggested solutions, bits of new information can be provided.              \\ \midrule
Decision-Making & --                      & The proposals of the argumentative phase are formulated into concrete action steps.   \\ \midrule
Concluding      & Appreciative Reflection & Recapitulate the conversation benevolently.                                           \\ \cmidrule(l){2-3} 
                & Appointment             & Make an appointment.                                                                  \\ \cmidrule(l){2-3} 
                & Farewell                & The interlocutors say their goodbyes.                                                 \\ \bottomrule
\end{tabularx}
\end{table}

\begin{table}
\caption{Communication techniques, their subcategories and short descriptions following \citet{bay2021erfolgreiche, gerich2016teachers,benien2003schwierige} and \citet{gartmeier2018gesprache}.}
\label{tab:techniques}
\begin{tabularx}{\linewidth}{@{}p{1.5cm}p{1.7cm}X@{}}
\toprule
\textbf{Technique} & \textbf{Subtechnique}        & \textbf{Description}                                                                                                                                                                                                                                                  \\ \midrule
Verbalising        & Undefined Attention Reaction & Short verbal signals, repetition of key words, prompt to say something.                                                                                                                                                                                               \\ \cmidrule(l){2-3} 
                   & Statement                    & Directly naming a feeling.                                                                                                                                                                                                                                            \\ \cmidrule(l){2-3} 
                   & Clarifying Question          & Clarification of a feeling with vague hints.                                                                                                                                                                                                                          \\ \cmidrule(l){2-3} 
                   & Further Question             & In-depth emotional statement, open question.                                                                                                                                                                                                                          \\ \midrule
Paraphrasing       & --                           & The key message of what has been said is repeated in the speaker's own words. The other person's statements are reproduced in their own words and capped if necessary.                                                                                                \\ \midrule
Structuring        & --                           & The teacher switches to a metacommunicative level in order to control the timing and/or content of the conversation and make it transparent. The structuring points to the following conversation or refers to the previous conversation. Key aspects are summarized. \\ \bottomrule
\end{tabularx}
\end{table}

For paraverbal communication we use the duration of the conversation in seconds, the duration of segments in seconds, the amount of words in a segment, the average length of words, the speaking rate in words per second, the shares of statements and questions, the textual sentiment, and the voice's pitch and loudness.

For nonverbal communication we use subsets of mimic, i.e. gaze and mutual gaze, smiling and mutual smiling, and some facial expressions, i.e. happiness, sadness, and anger. \\

Annotation of phases and techniques is done in {INCE}p{TION} (\citet{klie-etal-2018-inception}) by students and experts. This data will be used for (I3).
Analysis and visualization of paraverbal and nonverbal features is done using NOVA (\citet{Baur2020}) and its backbone DISCOVER (\citet{schiller2024discover}) on the recordings of the simulated conversations. The results are used to address (I1), and also shown to the students as part of the AI-based feedback and interview to address (I2).

\subsection{Feature Extraction}

\begin{figure*}[h]
  \centering
  \includegraphics[width=0.9\linewidth]{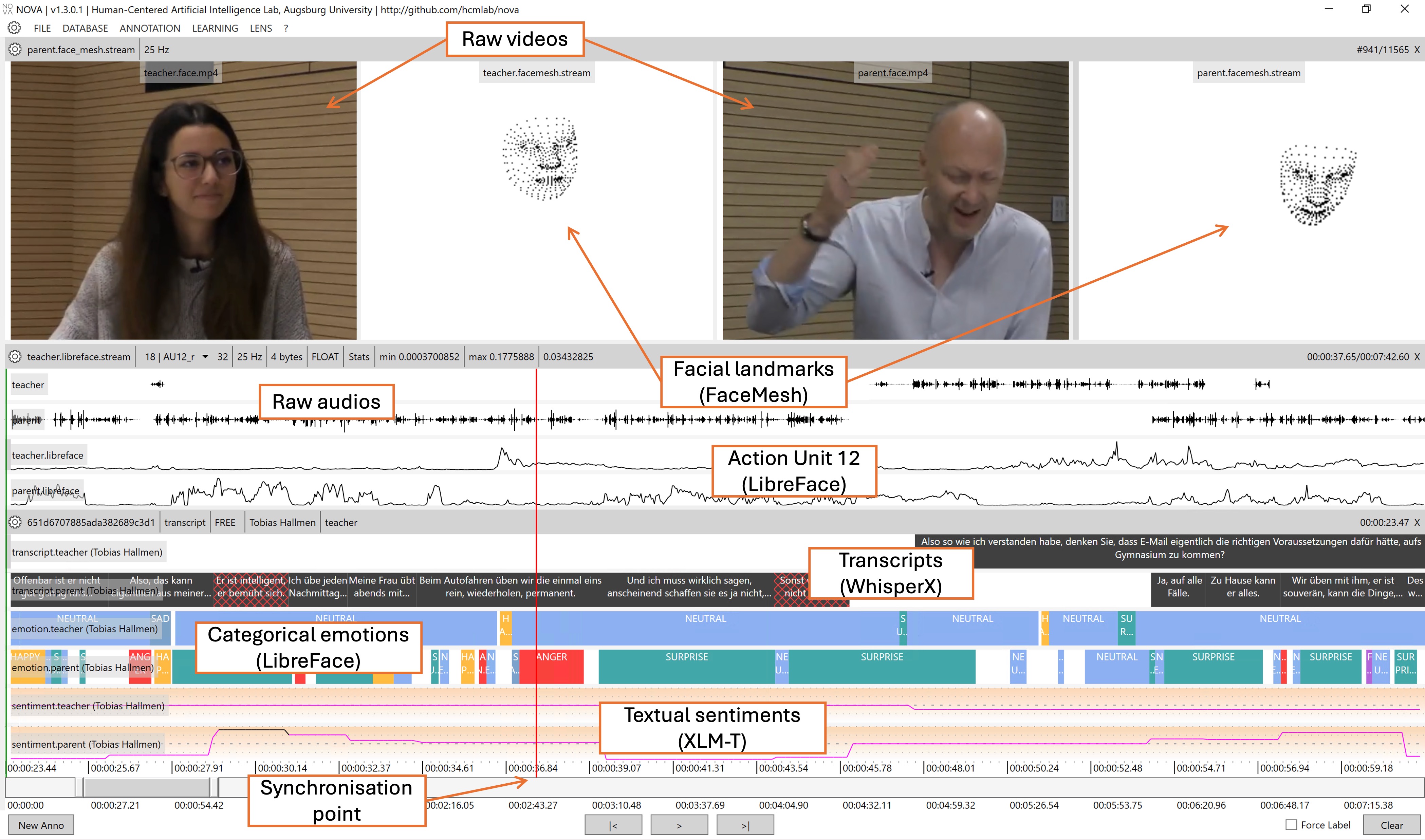}
  \caption{Exemplary picture of a simulated counselling in NOVA.}
  \label{fig:nova}
\end{figure*}

The recorded simulations are imported into NOVA, a tool used for exploratory data analysis, annotation, and extraction and visualization of machine learning and AI-based features. Figure~\ref{fig:nova} shows a screenshot of  the visualization in NOVA, which is modular and can be changed depending on the needs of the user or use case. The raw video streams are shown at the top, and audio streams are shown in the middle part as top horizontal tier.

Further streams can be extracted and shown -- in this example 3d facial landmarks in the top using the model FaceMesh by \citet{kartynnik2019facemesh}, and based on these LibreFace by \citet{chang2023libreface} is applied, yielding Action Unit 12 (AU12, ``lip corner puller'', \citet{ekman1976facs}) and categorical emotions as horizontal tiers. Based on the audio, a transcription and alignment of transcribed segments in time is done using large-v3 from WhisperX by \citet{bain2022whisperx}. These text transcriptions are then rated for sentiment using XLM-T by \citet{barbieri2022sentiment}. All these different kind of streams are synchronized and can be played back together -- the red vertical bar marks the current point in the timeline at the bottom.

Additional features, that have been extracted and are used in preparing AI-based feedback but not shown in the screenshot, are the gaze angle regression of OpenFace by \citet{baltruvsaitis2016openface}, face detection with BlazeFaze from \citet{bazarevsky2019blazeface}, and audio functionals provided by openSMILE from \citet{eyben2010opensmile}. 

Own models were developed on different datasets for detecting smiling and gazing in the ``main direction'', i.e. the interlocutor. They take the regressed Action Units, respectively gaze angles, as input.

\subsection{Preparing AI-based Feedback}
\label{sec:prep-ai-feedback}
In fusing all the aforementioned features together, the transcription by WhisperX plays a central role: not only does it enable the annotation of verbal skills, i.e. conversational phases and communication techniques, cf. Section~\ref{sec:background}, but by providing timestamps for the transcribed segment, it enables splitting the simulation into time intervals of speaking and listening. This distinguishing is crucial for computing the feedback for paraverbal (spoken) and nonverbal (spoken and unspoken) features. Currently, the verbal features are only collected by having the students annotate their own and two other sessions' transcript as a form of self-reflection and deepening of the learned theories.

We segment the transcribed text using voice activity detection (VAD) in WhisperX. It tries to segment on sentence level, but depending on speaking rate, loudness, and bigger pauses while speaking, thereby influencing VAD, it can happen that a segment contains multiple sentences -- rarely an incomplete sentence. 
We split segments containing several sentences according to punctuation. The incomplete sentences can not be handled automatically and lead to slight errors downstream. 

To make best use of the data at hand, the video recordings define the time granularity with $25fps$, i.e. everything is computed in frames of size $40ms$ before aggregating to session level statistics. Therefore the transcript with its timestamps is left joined to the frames, creating a distinction of frames with speech and without.

On the spoken frames, we compute the paraverbal features as follows: the session duration is the time difference in seconds of the last and the first spoken frame, and the segment duration is analogously the time difference of the last and the first spoken frame of the respective segment. The word count per segment is computed by counting white spaces plus one. The word length is computed as the average of characters per word in a segment, serving as a proxy for complexity of the language used. The speaking rate is defined as the quotient of words per segment and segment duration. The share of statements and questions are calculated by counting the punctuations of sentences, i.e. full stops and question marks. The textual sentiment, ranging from negative ($-1$) over neutral ($0$) to positive ($+1$) is computed by feeding the segments into XLM-T and aligning the output with the respective spoken frames. Similar is done for pitch ($Hz$) and loudness (based on auditory model), here we align the audio-based output of openSMILE with the respective spoken frames, i.e. the current spoken frame is extended with three preceeding frames, yielding a window size of $200ms$ containing $200ms * 16kHz = 3200$ audio samples.

Since nonverbal communication happens all the time, we compute the nonverbal features frame-wise over the whole session: for gaze, meaning mostly but not exclusively eye contact, we collect OpenFace's gaze angles against x- and y-axis and cluster these agglomeratively into ``main direction'' and ``other''. The assumption here is that the main direction resembles either the interlocuter, or directions which are part of the current speech, e.g. showing the parent a brochure and talking about the contained information. The hyperparameters were chosen in such a way on a different, private dataset, featuring counselling in medical doctor and patient conversations, that the amount of gaze in main direction correlated significantly with the standardized patients' satisfaction questionnaire (SPSQ, \citet{chessman2003assessing, choudhary2015teaching}). Mutual gaze is when both participants are gazing in the main direction.

The smile detector was trained on by OpenFace extracted Action Units in NoXi, a dataset by \citet{cafaro2017noxi} featuring interactions of experts and novices containing among others, annotations for smiling. Mutual smiling is when both interlocuters smile.

For categorical emotions we opted for the most typical ones in teacher-parent conversations according to our experts' experience: happiness, sadness, and anger. These are predicted by LibreFace after cropping and aligning the detected faces using BlazeFace and FaceMesh onto standardized faces.

All these paraverbal and nonverbal features are then aggregated on session level by averaging over segments, respectively the whole session, and presented to the students in different visualizations.

\subsection{Evaluation}
(I1) To quantize the predictive tendencies of features as seen in Section~\ref{sec:ai-based-feedback}, we conducted a five-fold stratified cross-validation using different sets of features as independent and expert rating as dependent variables. 

As classifiers we chose logistic regression (LR) by \citet{cox1958logisticregression}, Support Vector Classifier (SVC) by \citet{cortes1995svm}, and Extreme Gradient Boosting (XGB) by \citet{Chen_2016xgboost}. For feature sets we chose intuitively paraverbal, nonverbal, and paraverbal+nonverbal, as well as an computed selection by FeatureWiz (\citet{seshadri2020featurewiz}):  question, statement, sentiment, gaze, smile, happines, sadness, and anger.


XGB achieved highest average accuracy of $56.0\%$ and standard deviation of $16.0\%$ on the computed feature set by FeatureWiz. LR and SVC both achieve second highest accuracy with $55.3\%$ and standard deviation of $6.9\%$ on the feature sets nonverbal and FeatureWiz.

(I3) To evaluate language models on verbal skills we need to have a solid groundtruth first. 
Inter-rater agreement is 
$66.9\%$ for conversational phases and $72.4\%$ for communication techniques. 

\begin{figure}[h]
  \centering
  \includegraphics[width=0.75\linewidth]{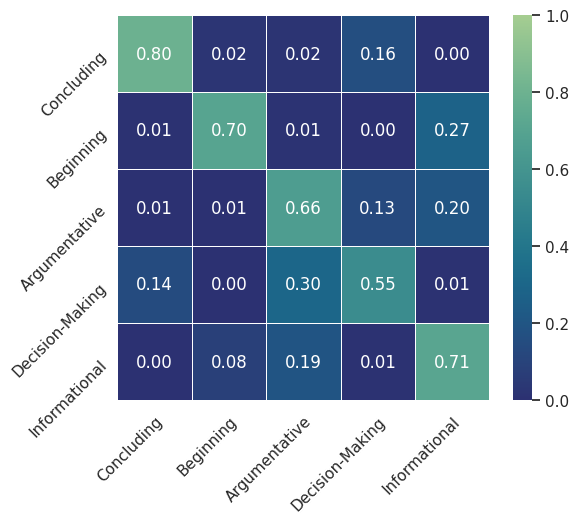}
  \caption{Coincidence matrix for annotation of conversational phases.}
  \label{fig:phase-coincidence}
\end{figure}

\begin{figure}[h]
  \centering
  \includegraphics[width=0.60\linewidth]{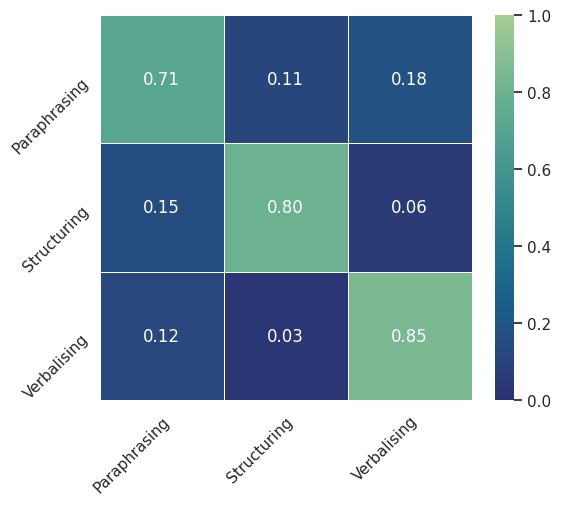}
  \caption{Coincidence matrix for annotation of communication techniques.}
  \label{fig:technique-coincidence}
\end{figure}

When looking at coincidence matrices in Figures~\ref{fig:phase-coincidence} and~\ref{fig:technique-coincidence}, we see $18\%$ of the technique paraphrasing was confused with verbalising and $15\%$ of structuring with paraphrasing. Looking at the phases, we see $27\%$ confusion between beginning and informational, $20\%$ between argumentative and informational, and $30\%$ between concluding and argumentative phases.

\section{Qualitative User Study}
\label{sec:study}

We conducted guideline-based interviews following \citet{helfferich2022leitfaden} and audio-recorded them. These are then transcribed according to instructions by \citet{dresing2015praxisbuch}. Integrating WhisperX into this process reduced time requirements by two-thirds. The transcriptions' content is then qualitatively analysed afterwards using the method described by \citet{kuckartz2012qualitative}. As the interviews are in German, we provide translations.

The students' first received the collective feedback in a group setting, followed by their individual feedback during the interviews. Questions could be posed at any time. The study is used to address (I2). 


\subsection{AI-based Feedback}
\label{sec:ai-based-feedback}
\begin{figure*}[h]
  \includegraphics[width=\linewidth]{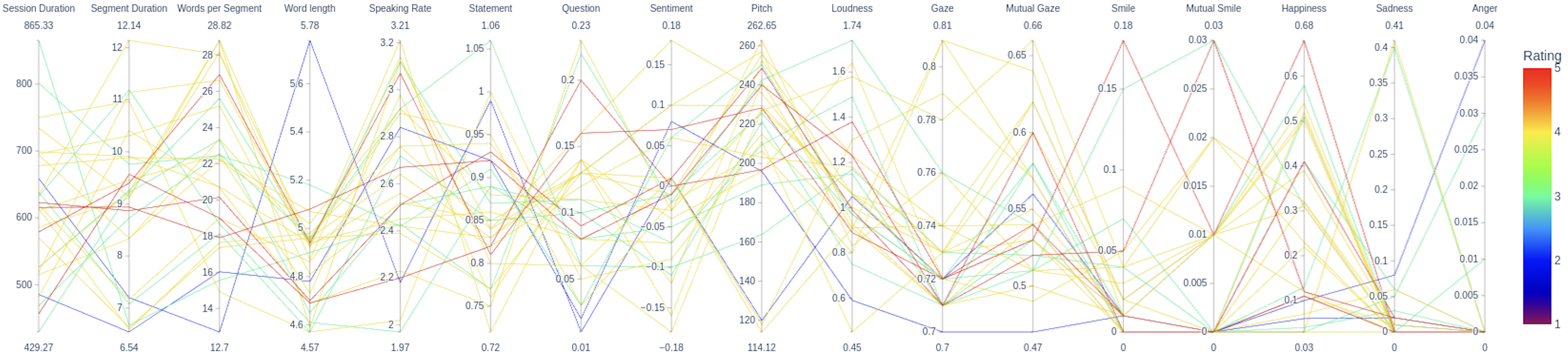}
  \caption{Collective feedback with each line resembling a student and showing the session averages on each axis.}
  \label{fig:collective-all}
\end{figure*}

\begin{figure}[h]
  \centering
  \includegraphics[width=0.75\linewidth]{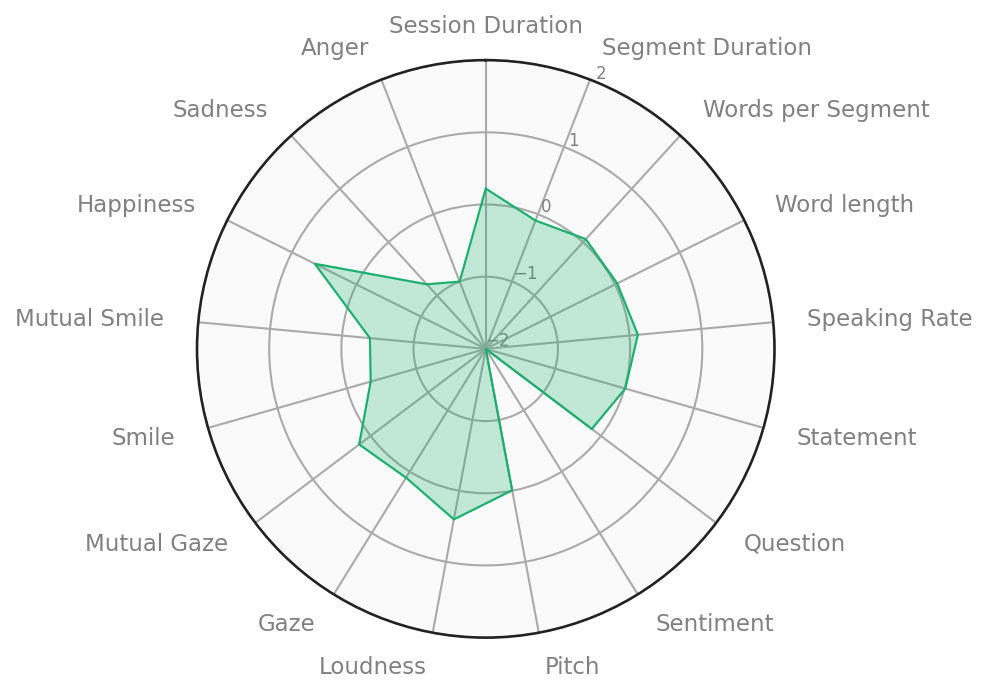}
  \caption{Exemplary radar chart showing individual feedback.}
  \label{fig:individual-radar}
\end{figure}

The AI-based feedback consists of multiple parts:

\begin{enumerate}[label=(\roman*)]
    \item tabular overview of session-aggregated features showing both absolute and deviations from the students' mean as relative numbers, cf. Table~\ref{tab:collective-feedback}
    \item parallel coordinate plots showing all or by rating grouped sessions, cf. Figure~\ref{fig:collective-all}
    \item radar charts per session, cf. Figure~\ref{fig:individual-radar}
    \item NOVA play backs per session, cf. Figure~\ref{fig:nova}
\end{enumerate}

These can also be understood as form of collective (i \& ii) and individual (iii \& iv) feedback, as well as containing an interpretation of the features (ii \& iii), i.e. rating by experts, or none, thereby being non-evaluating (i \& iv). For rating the counselling in simulations, the experts used a five-point Likert scale with 1 = ``not helpful at all'' and 5 = ``very helpful''.

\paragraph{(i) Tabular Data}

In Table~\ref{tab:collective-feedback} is an excerpt listing only two instead of all students showing the session-aggregated features. For easier peer-comparison the table also includes deviations from the mean of all students in multiples of the mean as relative values:

\begin{equation*}
    relative = \frac{absolute-mean}{mean} = \frac{absolute}{mean} - 1
\end{equation*}

For example, student A has a session duration of $693.96s$, which is $23\%$ greater than the mean ($= 564.20s$). This definition introduces the special values $relative = 0$, i.e. $absolute = mean$ (e.g. student A's statement share), and $relative = -1$, i.e. $absolute = 0$ (e.g. student B's mutual smile share).

The table does not show any expert rating or any suggestive interpretations but only the raw numbers. This encourages students to theory-craft their own opinions on how to interpret the results and compare themselves to each other.

\begin{table}
\caption{Excerpt of the tabular data shown to students as part of collective feedback.}
\label{tab:collective-feedback}
\begin{tabular}{@{}lrrrr@{}}
\toprule
                  & \multicolumn{2}{c}{\textbf{Student A}} & \multicolumn{2}{c}{\textbf{Student B}} \\
\textbf{Feature}  & \textbf{absolute}  & \textbf{relative} & \textbf{absolute}  & \textbf{relative} \\ \midrule
Session Duration  & 693.96             & 0.23              & 658.80             & 0.16              \\
Segment Duration  & 8.34               & -0.05             & 7.20               & -0.18             \\
Words per Segment & 22.19              & 0.04              & 12.70              & -0.40             \\
Word length       & 5.02               & 0.02              & 5.78               & 0.18              \\
Speaking Rate     & 2.92               & 0.10              & 2.18               & -0.18             \\
Statement         & 0.88               & 0                 & 0.99               & 0.14              \\
Question          & 0.11               & 0.09              & 0.01               & -0.93             \\
Sentiment         & -0.03              & -0.83             & 0.01               & 1.42              \\
Pitch             & 226.84             & 0.08              & 119.99             & -0.43             \\
Loudness          & 1.58               & 0.41              & 1.05               & -0.06             \\
Gaze              & 0.78               & 0.07              & 0.72               & -0.02             \\
Mutual Gaze       & 0.66               & 0.20              & 0.56               & 0.02              \\
Smile             & 0.04               & 0.35              & 0.01               & -0.79             \\
Mutual Smile      & 0.01               & 0.08              & 0                  & -1                \\
Happiness         & 0.54               & 0.83              & 0.10               & -0.67             \\
Sadness           & 0                  & -1                & 0.08               & 0.62              \\
Anger             & 0                  & -1                & 0.04               & 9.45              \\ \bottomrule
\end{tabular}
\end{table}

\paragraph{(ii) Parallel Coordinates}
The parallel coordinates plot (Figure~\ref{fig:collective-all}) 
was shown both in per session and in sessions grouped by rating form. 
Each axis is one feature and every line resembles one simulation, i.e. one student. The range of the axes is minimum to maximum of observed values, thereby visualising the spread and variance in each feature. The colour is defined by the experts' rating, i.e. a line in blue is rated as ``not very helpful'' ($= 2$) and one in red is deemed ``very helpful'' ($= 5$).

This makes it easy to spot outliers, e.g. max value in word length, the skewedness of the distribution, e.g. left-skewed for sentiment, and the kurtosis, e.g. happiness has bigger tails. Note the maximum value of $106\%$ of statements for one session -- this is caused by difficulties in VAD of WhisperX, cf. Section~\ref{sec:prep-ai-feedback}.

Tendencies you can see in the all-session graphs are enhanced when grouping sessions by rating: here you can give general recommendations to the collective of students on how to improve if you orientate along the best rated performances, i.e. the red lines. 
Looking at the features one advice could be: keep it short, be elaborate in your utterances, do not use complex words, speak at a normal rate, do ask questions, be neutral to slightly positive in your sayings, and emphasize with volume.

Keep eye contact, but do not stare, smile and if possible, smile mutually, display a happy face and avoid making a sad or angry face.

\paragraph{(iii) Radar Charts}
The individual radar charts, cf. Figure~\ref{fig:individual-radar}, are a combination of the approaches in (i) and (ii): compare each student and every paraverbal and nonverbal feature to the means of the group of best rated students and show deviations in multiples of the mean. So ideally, a ``very good'' student has a circle along the $0$-line. To keep the graphic clear, the deviations are clipped to $-2$ and $2$.

Taking a look at an exemplary radar chart of a student in Figure~\ref{fig:individual-radar}, an individual recommendation for action could be: the overall duration, your used language complexity, how fast and how long you speak during your turns, the amount of questions and statements you make are very good. Keep your gaze, loudness, and happy face. Express yourself less negatively, try to smile a bit more and dare to show sadness and anger if it fits the situation.

\paragraph{(iv) NOVA Play Back}
This individual feedback is highly interactive as the student's recorded simulation is loaded into NOVA for play back and enriched by a selection of extracted features 
, cf. Figure~\ref{fig:nova}. Here the students could choose from simply playing back a couple of minutes from the beginning, or start with a scene they remember as crucial, or search for scenes with possible misclassifications, e.g. a student denies making angry-looking faces and is convinced after showing a couple of scenes classified as ``anger''. This NOVA play back serves as a stimulus for the interview, as self-reflection, and as a means to rule out wrong self-perception.

\subsection{Students' Feedback}
With regard to the initial impression of NOVA, the students showed great interest in the AI tool, which was perceived by many as functional. Questions about NOVA annotation were asked, such as: ``What exactly is this teacher audio or parent audio again?'' (W1-02, pos. 72) and the advantages of the tool were identified: ``So for the nonverbal I found it good.'' (W1-03, pos. 30).

The induced comparison  of NOVA playback (iv), the tabular data (i), the parallel lines (ii),  and the radar charts (iii) revealed a differentiated picture. Many of the students stated that (iv) was preferable. This decision was justified on the one hand by the possibility of being able to better observe the progression of the relevant aspects in the interview in (iv). On the other hand, the validity of (iv) played a role: ``I prefer the video excerpts [...] because then you have a visual representation of it and can see it right away: Why did NOVA now display this? [...] If you only have the facts, then it is often so implausible.'' (W1-01, pos. 24).

Other participants preferred (i). This was justified with the better clarity of the presented numbers: ``Because it was easier for me to read.'' (W3-03, pos. 115). In contrast, (iv) was perceived as more difficult to access: ``With the video, it is more difficult to watch everything at the same time'' (W3-02, pos. 48). In addition, it was highlighted that only (i) offers a way to classify these characteristics in relation to other students. ``I think it's the best way to read it. You also have the comparison and I find that exciting.'' (W4-03, pos. 221)  Some students saw the combination of (i) and (iv) as the optimal solution: ``So in the video you can see the situation again directly. In the data sheet [i] you have the overall average and also the comparison with the others.'' (W1-06, pos. 88).

Finally, in comparing Radar Charts (iii) to the former to forms of feedback (iv \& i), some students needed a deeper explanation of their functionality, since they initially interpreted it just as a different presentation of the data in (i): ``I guess it also reflects the list.''  (W4-01, pos. 174). However, once the added value was explained again, the participants were able to understand it: ``Apparently I am beeing perceived as more positively than i thought.'' (W4-04, pos. 215). Some participants preferred this form of feedback over the other two, since deviations from the optimum were best visible here: ``Just take a look to see if anything stands out—maybe there are a few areas you could work on.'' (W4-03, pos. 268). 

There were many suggestions regarding the design perspectives for NOVA. First of all, desired additional features were identified. Students mentioned obvious aspects such as voice variance, but also suggested more far-reaching ideas, such as the integration of posture or vital data: ``A dream thing would be [...] to be able to measure heart and breathing rate to notice when someone is [...] upset'' (W1-04, pos. 53). Suggestions were also made for the further development of existing features. For example, there was a desire to determine emotions by means of voice variance and to display the sentiment values as a progression in order to make the reciprocal influences of the participants in the conversation visible: ``Maybe implementing the parent's mood curve in the course of the conversation, so that you can spot an entrancement in the end?'' (W2-02, pos. 74). 

Finally, the need for more interpretation of the data was expressed again: ``It would be interesting to see whether the counselling is coherent in itself, but the quality of the counselling bits would have to be assessed.'' (W2-07, pos. 81).

For the NOVA video interface, the lines that visualized the characteristics were partly perceived as helpful, but their number was overwhelming: ``There were a lot of lines and either you just concentrated on one, otherwise it didn't work at all.'' (W1-05, pos. 24). On this note one student suggested to highlight certain parts of the NOVA-feed: ``If it would light up green when I smiled, it might be more visual. [...] Maybe even highlight special peaks or something.'' (W3-07, pos. 102 - 104). 


\section{Discussion}
Returning to the guiding investigations previously posed, we will now address each of them in turn:

(I1) Due to the small sample size we see four times the same result when correlating features to experts' ratings despite different classifiers and feature sets. Nevertheless we do see predictive capabilities of features for the rating ($56.0\%$) above dummy classification level, which is $50\%$ across folds for the majority class (4 = ``helpful''). With more data collected every semester as part of future work, the patterns may become clearer. This would also help the diversity of the data as well as the generalizability of findings.

Regarding (I2) it can be concluded that AI-based Feedback as a whole was positively perceived by the students. Especially the combined presentation of NOVA playback and the tabular data appeared to be promising. Furthermore the Radar Charts could support students by identifying areas on which they could work on. The AI-based Feedback was furthermore able to awaken a thirst for data in the participating students. Vital data and gradient curves were identified as particularly interesting. 
Ultimately, an interpretation and prioritization of the data still appeared to be absent in the eyes of the students. Focusing on ‘golden moments’, as already emphasized in the work of (\citet{schiller2024discover}), may be helpful in identifying such tipping points within the conversations and, based on these, potential levers for adjustments. 

(I3) The inter-rater agreements for conversational phases ($66.9\%$) and communication techniques ($72.4\%$) are promising, but to be deemed very good, they have to be above $80\%$. This is corroborated by entries in the coincidence matrices that are above what could be human error (e.g. $5\%$). The possible reasons of these confusions have to be investigated: the annotation guidelines could be imprecise, the annotators understand the guidelines differently, or the classes are in fact not disjunct.



Therefore, to include verbal features as independent variables for predicting expert ratings as future work, we have to robustize the annotation guidelines first and then train a model to automate verbal feature extraction. These can then be used for AI-based feedback as well.

These findings suggest that AI-based feedback can significantly enhance teacher training programs by providing objective, actionable insights. However, the study's limitations, such as the small sample size and the need for more robust verbal feature annotations, must be addressed in future research.




\section{Conclusion}
In this work we have shown, that NOVA with its feature extraction and visualization capabilites works to provide AI-based feedback in simulations of counselling with prospective teachers, especially for paraverbal and nonverbal expression. The results in annotating verbal expressions are promising, 
there is still headroom for improvements.
In the current study, an AI expert was in the loop of the feedback process, helping to interpret parts of the data.
Future work will address providing such explanations automatically.
Overall, our study has demonstrated high potential of the collective and individual AI-based feedback for enriching teaching methods for learning skills related to teacher-parent conversations. Our method supports experts in giving objective feedback to students.

\section*{Safe and Responsible Innovation Statement}
This research adheres to ethical standards and prioritizes the well-being of all participants. AI-based feedback tools were developed and implemented with careful consideration of privacy and data security. The study ensures that all data is anonymized and used solely for educational purposes when presented to students, unless they gave consent, e.g. showing a recording or images in publications. The AI models employed are designed to provide constructive feedback without causing harm or distress. We are committed to continuous improvement and transparency, ensuring that our innovations contribute positively to the field of education while respecting the rights and dignity of all individuals involved.

\begin{acks}
This work was partially funded by the KodiLL project (FBM2020, Stiftung Innovation in der Hochschullehre).

AI tools were used to rephrase, translate, and perform spelling and grammar checks.

\end{acks}

\bibliographystyle{ACM-Reference-Format}
\bibliography{main}

\end{document}